\documentclass[aps,twocolumn,superscriptaddress]{revtex4-1}
\usepackage{graphicx}
\usepackage{dcolumn}
\usepackage{bm}
\usepackage{caption,subcaption} 
\usepackage{booktabs,array}
\usepackage{graphicx,caption,subcaption}
\usepackage{amsfonts}
\usepackage{amsmath,amssymb}
\usepackage{extarrows}
\usepackage{color}
\usepackage{slashed}
\usepackage{epstopdf}
\usepackage{esvect}
\usepackage[english]{babel}

\begin{document}

\title{The Pauli Form Factor of Quark and Nontrivial Topological Structure of the QCD}%
\author{Baiyang Zhang}%
\email{zhangbaiyang@impcas.ac.cn}
 \affiliation{
Institute of Modern Physics, Chinese Academy of Sciences, Lanzhou 730000, China}\affiliation{
University of Chinese Academy of Sciences, Beijing 100049, China
}

\author{Andrey Radzhabov}%
 \email{aradzh@icc.ru}
 \affiliation{
Institute of Modern Physics, Chinese Academy of Sciences, Lanzhou 730000, China} \affiliation{
Matrosov Institute for System Dynamics and Control Theory SB RAS Lermontov str., 134, 664033, Irkutsk, Russia
 }

\author{Nikolai Kochelev}
 \email{kochelev@theor.jinr.ru}
 \affiliation{
Institute of Modern Physics, Chinese Academy of Sciences, Lanzhou 730000, China} \affiliation{Bogoliubov Laboratory of Theoretical Physics, Joint
Institute for Nuclear Research,\\ Dubna, Moscow Region, 141980
Russia}

\author{Pengming Zhang}
\email{zhpm@impcas.ac.cn}
 \affiliation{
Institute of Modern Physics, Chinese Academy of Sciences, Lanzhou 730000, China}

\date{\today}

\begin{abstract}
We calculate the electromagnetic Pauli form factor of quark induced by the nontrivial topological fluctuations of QCD vacuum called instantons.
It is shown that such contribution is significant. We discuss the possible implications of our result in the photon-hadron reactions and in
the dynamics of quark-photon interactions in the dense/hot quark matter.

\end{abstract}

\pacs{ {13.40.Gp}, {12.38.-t}, {12.38.Lg}}
\keywords{quarks, form factor, instanton, quantum chromodynamics}

\maketitle

\section{\label{sec:intro}Introduction}
Nowadays the study of electromagnetic structure of the elementary particles is one of the hottest topics in the Standard Model (SM).
One well known puzzle is the experimental value of the muon anomalous magnetic moment which shows the significant deviation from the SM prediction (see recent reviews~\cite{Lindner:2016bgg,Dorokhov:2016knu}). Electromagnetic probes of the hadrons give very important information about the structure of the strong interaction \cite{Punjabi:2015bba}.
Various models have been developed which enable us to study electromagnetic properties of hadrons in terms of form factors~\cite{Altarelli:1995,Petronzio:2003,Scopetta:2003et,Gutsche:2014zua,Kochelev:2015pqd,Arrington:2006zm,Faccioli:2002jd,Faccioli:2002wr,Faccioli:2001qg,Blotz:1996ad}.
%
In the past decades the significant progress has been made, especially regarding the study of the relation between generalized parton distribution functions (GPD) and electromagnetic form factors of hadrons~\cite{Ji:1996,Radyushkin:1996nd}.
The quark form factors carry the information about internal structure of the constituent quark and provide the bridge between partonic picture of the hadrons and their constituent structure
\cite{Kochelev:2015pqd,kochelev1,Kochelev:2003,Dorokhov:2004fb,Petronzio:2003,Simula:2003,Roberts:2007ji}. \\
In this paper we consider a new nonperturbative contribution to electromagnetic Pauli form factor (EPFF) of quark arisen from instanton induced quark-gluon vertex.
 The instanton is the well-known solution of QCD equation of motion in the Euclidian space-time which
has nonzero topological charge. It was
shown that instantons play a very important role in hadron physics (see the reviews
\cite{Schafer:1998, Diakonov:2002,Kochelev:2005xn}).
In particular, the instantons leads to the spontaneous chiral symmetry breaking (SCSB) in
strong interaction which is not only one of the main sources for the observed hadron masses but also
leads to the various anomalies observed in the
spin-dependent cross sections~\cite{kochelev1,Ostrovsky:2004pd,Cherednikov:2006zn,Hoyer:2005ev,Kochelev:2013zoa,Qian:2015wyq,Kochelev:2015pqd}.
One of the cornerstones of the instanton-based theory of the spin effects in the strong interaction is the instanton-induced anomalous chromomagnetic
quark-gluon interaction introduced in \cite{kochelev1}. The strength of this interaction is determined by the
dynamical mass of the quark in the instanton vacuum \cite{Diakonov:2002,kochelev2} which is
directly related to the phenomenon of the SCSB. The first attempt to estimate the effect of instantons to EPFF was made in \cite{Kochelev:2003} where
the so-called instanton's perturbative theory was used. This approach was developed in the papers
\cite{ringwald1,ringwald} to obtain the effect of the small size of the instantons to the Deep-Inelastic Scattering (DIS) at large
transfer momentum $Q^2=-q^2$. However, the final result for their contribution to DIS at the large $Q^2$ was found to be very small.
The same conclusion is also valid for the contribution of the small instantons to the large $Q^2$ asymptotic of the EPFF of quark
obtained in \cite{Kochelev:2003}. Here, we will use another way to calculate the instanton contribution to EPFF. This approach is based on the
effective quark-gluon vertex induced by instantons and allows to obtain the prediction for EPFF in the wide interval of the $Q^2$ including even
very important case of the real photon, $Q^2=0$.

{\bf }

\section{The contribution of anomalous quark-gluon interaction to the quark electromagnetic form factor}

 The general vertex for photon-quark interaction for on-shell quark is
 \begin{align}
 \label{eq:qed}
 \Gamma^\mu = \gamma^\mu F_1(Q^2) + \frac{i\sigma^{\mu\nu}q_\nu}{2 M_q}F_2(Q^2)
 \end{align}
 where $F_1,F_2$ are electromagnetic Dirac and Pauli form factors, respectively, $M_q$ is the dynamical mass of the quark and
 $\sigma_{\mu\nu}= i(\gamma_\mu\gamma_\nu - \gamma_\nu\gamma_\mu)/2$.
 The anomalous quark-gluon chromomagnetic (AQGC) vertex induced by the instantons can be written in the form ~\cite{kochelev1,Diakonov:2002,kochelev2}
 \begin{equation}
V^a_\mu(k_1^2,k_2^2,t^2)=
\frac{ig_s\sigma^{\mu\nu}q_\nu}{2M_q}F_2(k_1^2,k_2^2,t^2)t^a,
 \label{vertex}
 \end{equation}
 where $k_1^2$ and $k_2^2$ are the virtuality of the initial and final quarks, respectively, $t=k_1-k_2$
 and the general case for non-zero virtualities of quarks and gluon is considered.
 The form factor $F_2(k_1^2,k_2^2,t^2)$
 suppresses the AQGC vertex
at short distances when the respective virtualities are large. Within the
instanton model it is explicitly related to the Fourier-transformed of both quark
zero-mode and instanton field, which take the forms
\begin{equation}
 F_2(k_1^2,{k_2^\prime}^2,t^2) =\mu_a
F_q(|k_1|\rho/2)F_q(|k_2|\rho/2)F_g(|t|\rho) \ , \label{FF}
\end{equation}
where
\begin{eqnarray}
F_q(z)&=&-z\frac{d}{dz}(I_0(z)K_0(z)-I_1(z)K_1(z))\nonumber \\
F_g(z)&=&\frac{4}{z^2}-2K_2(z), \label{ffg}
\end{eqnarray}
$I_{\nu}(z)$, $K_{\nu}(z)$ are the modified Bessel functions,
$\rho$ is the instanton size and $\mu_a=F_2(0,0,0)$ is the anomalous quark chromomagnetic moment (AQCM).
Within the instanton liquid model \cite{Schafer:1998}, \cite{Diakonov:2002},
where all instantons have the same size $\rho_c\approx 1/3$ fm, AQCM is
\cite{kochelev2,Diakonov:2002}
\begin{equation}
\mu_a=-\frac{3\pi (M_q\rho_c)^2}{4\alpha_s(\rho_c)}.
\label{AQCM}
\end{equation}

\begin{figure}[htbp]
 \begin{subfigure}[b]{0.5\linewidth}
 \centering
 \includegraphics[width=\linewidth]{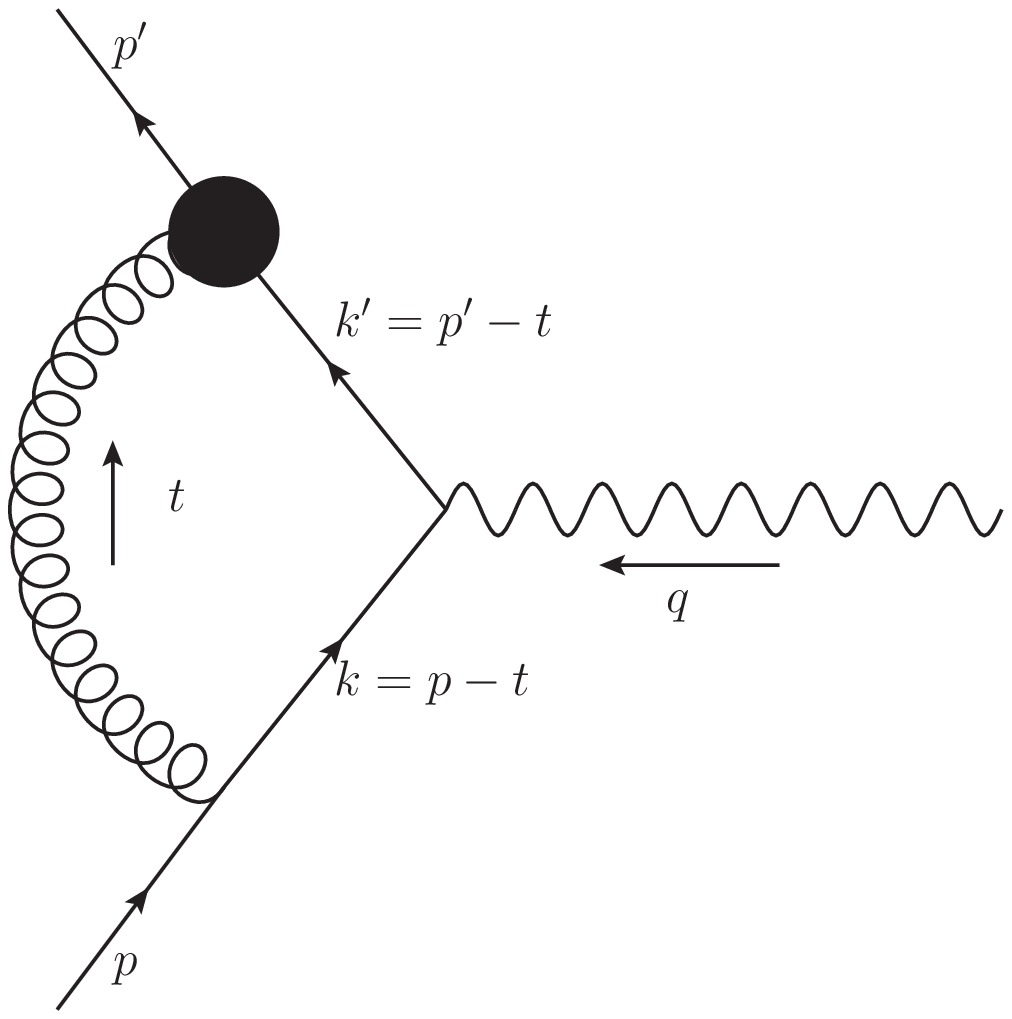}
 \caption{}\label{fig:am}
 \end{subfigure}%
 \begin{subfigure}[b]{0.5\linewidth}
 \centering
 \includegraphics[width=\linewidth]{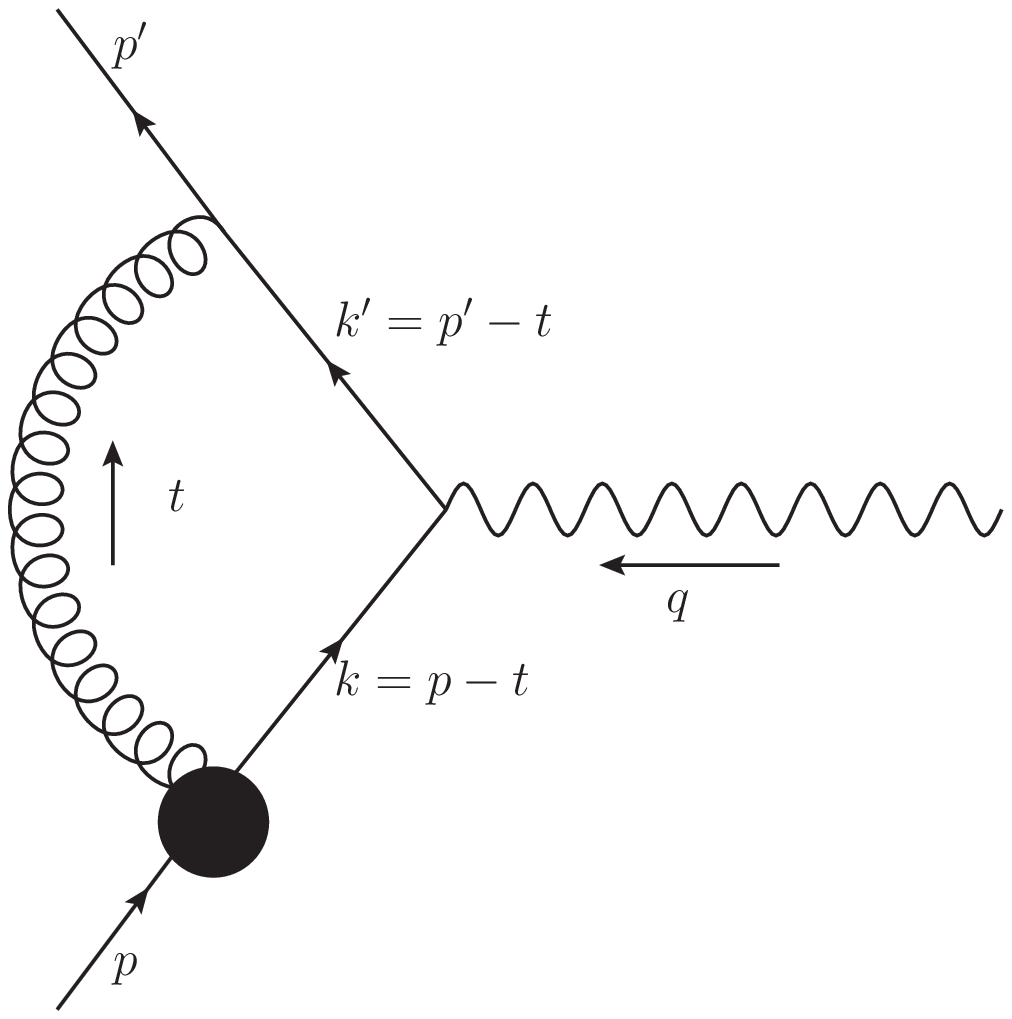}
 \caption{}\label{fig:am2}
 \end{subfigure}%
 \caption{\raggedright The diagrams with anomalous quark-gluon chromomagnetic vertex induced by instantons which contribute to
 EMFF of the quark.
 The vertex is denoted by a solid blob.}\label{fig:fd}
 \end{figure}

The first feature is that the strong coupling constant enters into the
denominator showing a clear nonperturbative origin of AQCM. The
second feature is the negative sign of AQCM. As we will see below,
this sign of AQCM leads to the positive sign of the anomalous quark magnetic moment (AQMM). The value of AQCM strongly depends on the
dynamical quark mass which is $M_q=170$ MeV in the mean field
approximation (MFA)~\cite{Schafer:1998}
 and $M_q=350$ MeV in the Diakonov-Petrov model (DP)~\cite{Diakonov:2002}.
Therefore, for the value of the strong coupling constant in
the instanton model, $\alpha_s(\rho_c) \approx 0.5$ and average size of instantons $\rho_c=1/600$ MeV$^{-1}$ \cite{Diakonov:2002} we get
\begin{equation}
{\mu_a}^{\mathrm{MFA}}=-0.4 \ \ \ \mu_a^{\mathrm{DP}}= -1.6
\label{mu}
\end{equation}
 The contribution to the electromagnetic Pauli form factor coming from the AQGC vertex is obtained by the consideration of the diagrams presented
 in Fig.~\ref{fig:fd}.

 To perform analytical calculations the gaussian approximation for the form factors in Eq.(\ref{ffg})
 \begin{equation}
 F_g(k_E^2) \approx F_q(k_E^2)\approx e^{-k_E^2/\Lambda^2},
 \label{approx}
 \end{equation}
 is used with $\Lambda=2/\rho_c$.

 The contribution from Fig.~\ref{fig:fd}(\subref{fig:am2}) is the same as that from Fig.~\ref{fig:fd}(\subref{fig:am}), hence the final result should be doubled. Therefore the total matrix element is
 \begin{align}
 \notag
 i\mathcal{M} & \equiv -e_qC_F g_s^2 \frac{\mu_a}{M_q} \int \frac{d^4 t}{(2\pi)^4} \frac{F_g(t^2) F_q({k^\prime}^2)N}{({k^\prime}^2 - M_q^2)(k^2-M_q^2)t^2} \\
 & = -ie_q\overline{u}(p')\Gamma^{\mu}(p,p') u(p),
 \label{eq:m}
 \end{align}
 where $C_F = \text{tr}(T^aT^a) = \frac{4}{3}$ is the color factor and $e_q$ is the electric charge of the quark and
 \begin{equation}
 N \equiv -i\overline{u}(p') \sigma^{\alpha\rho} (\slashed{k}' +M_q) \gamma^\mu (\slashed{k}+M_q) \gamma_\rho t_\alpha u(p).
 \label{eq:N}
 \end{equation}
 One way to extract Pauli form factor $F_2(Q^2)$ from $i\mathcal{M}$ is to rearrange the gamma matrices in Eq.~(\ref{eq:N}) and find the term proportional to ${i\sigma^{\mu\nu}}/{2M_q}$. However, a simpler way is to use projector operator method~\cite{Brodsky:1966mv,Knecht:2001}, by making use of identity
 \begin{equation}
 F_2(q^2) = \text{tr}\left\{ (\slashed{p}+M_q) \Lambda^{(2)}_\rho(p',p)(\slashed{p'}+M_q)\Gamma^\rho(p',p) \right\},
 \end{equation}
 where $q^2 = (p'-p)^2 \equiv -Q^2$ and
 \begin{equation}
 \Lambda^{(2)}_\rho(p',p) \equiv \frac{M_q^2}{k^2(4M_q^2-k^2)} \left[ \gamma_\rho + \frac{k^2+2M_q^2}{M_q(k^2-4M_q^2)}(p'+p)_\rho \right]. \nonumber
 \end{equation}
 By working in the Euclidean space-time,with the help of Feynman parametrization and identity
 \begin{equation}
 \frac{1}{k^n} = \int_0^\infty d\alpha \frac{\alpha^{n-1}}{(n-1)!} e^{-\alpha k},
 \end{equation}
 we obtain
 \begin{widetext}
 \begin{align}
 \notag
 F_2(Q^2) & =
 \frac{\mu_a e_q g_s^2}{12\pi^2} \int d^3x \int_0^\infty
	d\alpha
	\frac{\alpha^2}{\Delta^2}
	\text{Exp}\left\{ -M_q^2 \left[ \Delta(v_1+v_2)^2- \frac{1}{\Lambda^2} \right] - Q^2 \Delta v_1 v_2 \right\} \\
 & \;\;\;\; \times \left\{ \frac{3v_1+6v_2-7}{\Delta}- Q^2v_1 v_2 (v_1+2v_2-3) -M_q^2(v_1+v_2)((v_1+v_2)(v_1+2v_2)-2v_2)\right\},
 \label{eq:fr}
 \end{align}
 \end{widetext}
 where $\int d^3x \equiv 2\int_0^1 dx_1\; dx_2\; dx_3\; \delta(1-x_1-x_2-x_3)$ and
 \begin{align}
 \Delta & \equiv \alpha+\frac{2}{\Lambda^2},\\
 v_1 & \equiv x_2\frac{\alpha }{\Delta}, \\
 v_2 & \equiv x_1\frac{\alpha}{\Delta}+\frac{1}{\Lambda^2\Delta}.
 \end{align}
 \section{Numerical results}
 In our model the form factor $F_2(Q^2)$ is proportional to the quark charge. Therefore, there is the relation between u- and
 d-quark form factors
 \begin{equation}
 F_2^d(Q^2)=-\frac{1}{2} F_2^u(Q^2).
 \label{charge}
 \end{equation}
 \begin{figure}
 \centering
 \includegraphics[width=\linewidth]{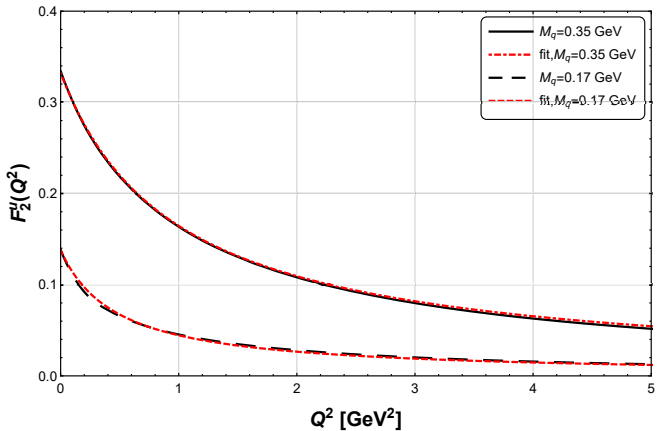}
 \caption{The $F_2$ form factor as a function of $Q^2$ for the different dynamical quark mass $M_q$ in the comparison with the fit
 given by Eq. (\ref{fit}).}
 \label{fig:f2Q2}
 \end{figure}

For simplicity, below only the result for u-quark case is presented in the figures.
In Fig.2 the result of the calculation of electromagnetic form factor as the function of $Q^2$ is presented for
two different masses of u-quark. Our numerical result can be fitted very well by the formula
\begin{equation}
F_2(Q^2,M_q)=\frac{F_2(0,M_q)}{1+\rho_cQ^2/(4.7M_q)},
\label{fit}
\end{equation}
which can be useful for the applications.
 We would like to emphasize that positive sign of $F_2$ form factor for u-quark (see Fig.2), is fixed by the negative
sign of the AQCM, Eq. (\ref{AQCM}). In the Fig.3 the dependency of the value of the magnetic moment of u-quark on the value of the
its dynamical mass is shown. It's behavior as a function of quark mass in the range between $80$ and $500$ MeV can be fitted
 very well with the linear function
 \begin{equation}
\mu_a^u\approx \frac{2}{3}\left (-0.065 + 0.97(M_q\rho_c)\right )\label{LowMassFit}.
\end{equation}

\begin{figure}[b]
 \centering
 \includegraphics[width=0.5\textwidth]{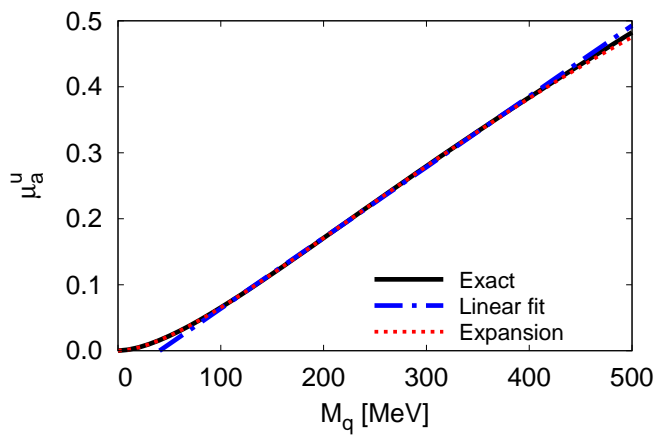}
 \label{fig:LowMassExp}
 \caption{Behavior of $\mu_a^u$ versus quark mass: exact expression (black solid line),
 linear fit (blue dashed-dotted), expansion Eq. (\ref{LowMassExp}) (red dotted). }
 \end{figure}

 The results for $\mu_a^{e,u}=F_2^u(Q^2=0)$ at the two different values of the dynamical quark masses obtained in
 mean field approximation ~\cite{Schafer:1998}
 and within the Diakonov-Petrov model~\cite{Diakonov:2002}
 are
\begin{eqnarray}
 \mu_a^{e,u}&=&0.33 \ \ \text{for} \ \ M_q=350 \, \text{MeV}, \ \ \nonumber \\
 \mu_a^{e,u}&=&0.14 \ \ \text{for} \ \ M_q=170 \, \text{MeV}.
 \label{value}
 \end{eqnarray}
 Our value for the quark magnetic moment at $ M_q=350$ MeV is in the qualitative agreement with the result of calculation within different approach based on the Dyson-Schwinger equation \cite{Chang:2010hb}. However, we would like to emphasize that in this paper the 
 $Q^2$ dependency of the EPFF is not considered.
 This $Q^2$ dependency in our model is presented in the Fig.2. One can mention its
 rather strong dependency on the virtuality of photon. In the model it is coming from the quark and gluon form factors presented in
 the Eq. (\ref{approx}).

\section{The large $Q^2$ behavior of EPFF}

The formula for expansion of the form factor at large $Q^2\gg M_q^2 $ is
\begin{align}
&F_2(Q^2)\approx -\frac{1}{Q^2}\frac{ e_q C_F g_s^2\mu_a }{4 \pi^2}
\sum\limits_{n=0}^{\infty} \int \limits_0^\infty dk^2 \frac{k^{2(n+2)}}{n!(n+2)!} M_q^{2n} \times
\nonumber\\
&\quad \times[F_gD_g]_{n+1}\left(2[F_qD_q]_{n-1}-M_q^2 \frac{k^{2}}{(n+3)}[F_qD_q]_{n+1} \right)\nonumber\label{F2MomentumAsymptotic},
\end{align}
where
\begin{align}
&[F_qD_q]_{-1} = -\int\limits_{k^2}^\infty dl^2 F_q(l^2)D_q(l^2) \nonumber \\
&[F_qD_q]_{0} = F_q(k^2)D_q(k^2) \nonumber \\
&[F_qD_q]_{+1} = -\left(\frac{d}{dk^2}\right) (F_q(k^2)D_q(k^2)) \nonumber \\
& ... \nonumber \\
&[F_qD_q]_{+n} = \left(-\frac{d}{dk^2}\right)^n (F_q(k^2)D_q(k^2)),
\end{align}
and the same formula for $[F_gD_g]_{i}$ with the corresponding changing of index
$q\rightarrow g$.
Using the expressions for the form factors in gluon and quark sector in Eq. (\ref{approx}) 
and $D_g(k^2)=1/k^2$, $D_q(k^2)=1/(k^2+M_q^2)$ one can rewrite it in the form
\begin{eqnarray}
&&F_2(Q^2)\approx\frac{1}{Q^2}\frac{ e_q C_F g_s^2\mu_a }{4 \pi^2} \biggl\{
\int \limits_0^\infty dk^2 \frac{e^{-k^2/\Lambda^2}}{k^2 + M_q^2}
 \nonumber \\
&&\quad \times \left[2 \Lambda^2 - e^{-k^2/\Lambda^2} (k^2 + 2 \Lambda^2)\nonumber \right]\nonumber \\
&&\quad
+\sum\limits_{n=1}^{\infty} \int \limits_0^\infty dk^2 \frac{k^{2(n+2)}M_q^{2n}}{(n-1)!(n+2)!} \label{asymp1}\\
&&\quad \times \left( \frac{2}{n}(F_gD_g)_{n+1}(F_qD_q)_{n-1}-(F_gD_g)_{n}(F_qD_q)_{n}\right)\biggr\}\nonumber
\end{eqnarray}
and show that the main contribution to EPFF is coming from the first term in brackets.
Moreover, one can perform additional expansion over $M_q^2/\Lambda^2$.
In this approximation and ignoring the second term in Eq. (\ref{asymp1}), one can write the leading orders in $M_q^2/\Lambda^2$ expansion in the form
\begin{eqnarray}
F_2(Q^2)&\approx&4e_q\frac{M_q^2}{Q^2}
\bigg( 2\ln(2)-\frac{1}{2}\nonumber \\ &+& \frac{M_q^2}{\Lambda^2} \left[\ln\left(8\frac{M_q^2}{\Lambda^2}\right)-2 + \gamma_E \right]\bigg).
\label{asymp3}
\end{eqnarray}
where
$\gamma_E$ is the Euler's constant.
By using the relation $\Lambda\approx 2/\rho_c$, it can be rewritten as
\begin{eqnarray}
F_2(Q^2)&\approx&4e_q\frac{M_q^2}{Q^2}
\bigg( 2\ln(2)-\frac{1}{2}\nonumber \\ &+& \frac{(M_q\rho_c)^2}{4} \left[\ln\left(2(M_q\rho_c)^2\right)-2 + \gamma_E \right]\bigg).
\label{asymp2}
\end{eqnarray}
Therefore, at large $Q^2$ form factor behaves as $F_2(Q^2)\sim 1/Q^2$.

\section{The low $Q^2$ behavior of EPFF}

It can be shown that in the limit $Q^2\rightarrow 0$ the form factor is
\begin{eqnarray}
F_2(0)&\approx &
e_q (M_q\rho_c)^2
\bigg( y + \frac{(192 y + 211)}{288}(M_q\rho_c)^2\nonumber\\
 &+& \frac{ (1536 y + 3089)}{9216}(M_q\rho_c)^4+...\bigg),
\label{LowMassExp}
\end{eqnarray}
where
\begin{equation}
y=\ln\left(\frac{2}{(M_q\rho_c)^2} \right)- \gamma_E - \frac{1}{4}\nonumber.
\end{equation}
The expansion given by Eq.\ref{LowMassExp} describes the exact result very well in the region of small $M_q\rho_c < 1$, Fig.3.
 One can see that $F_2(0)$ vanishes in the limit $M_q\rightarrow 0$.
 It means that this contribution to form factor is directly related to the phenomenon of SCSB.

\section{Conclusion}

In this paper, we calculate the quark electromagnetic form factor within the nonperturbative approach based on the instanton picture for the QCD vacuum.
It is shown that anomalous quark-gluon chromomagnetic interaction induced by instantons leads to large magnetic moment of u- and d-quarks. 
Possible applications of our results are as follows.
One of the tasks is
to consider the influence of EPFF on the  hadron electromagnetic form factors.  We would like to mention that
instanton contribution to the electromagnetic form factors of proton, neutron and pion were calculated using different versions of instanton model in \cite{Forkel:1994pf,Faccioli:2002jd,Faccioli:2002wr,Faccioli:2001qg,Blotz:1996ad} in semi-classical approaches to the corresponding
correlators.  However, it would be interesting to study the electromagnetic properties of hadrons based on the constituent quark model with an effective quark-photon and quark-gluon vertices  induced by instantons. In this way, one can take into consideration the confinement effects as well in spirit of the calculation of nucleon electromagnetic  form factors carried out in  \cite{Petronzio:2003}  for constituent quarks with inner structure.
It is evident that due to the existence of an additional scale in our model related to the instanton size $\rho_c\approx 1/3$ fm,
one can expect the deviation of the $Q^2$
dependency of the hadron form factors from the quark-counting rule prediction \cite{Matveev:1972gb,Brodsky:1973kr,Brodsky:1974vy}.

We should stress that EPFF leads to quark spin-flip. Therefore, it should make contribution to various
spin-dependent photon-hadron cross sections, including polarized semi-inclusive DIS.
Another possible application, in the line of 
\cite{Fayazbakhsh:2014mca}, is the study of the influence of the non-zero value of the
anomalous quark magnetic moment on the dynamics of Quark-Gluon Plasma (QGP) in the strong magnetic field.
We would like to emphasize that our new type of photon-quark interaction is very sensitive to the
topological structure of the QCD vacuum which might be drastically changed during the deconfinement transition \cite{Ilgenfritz}. This
 phenomenon can lead to, for example, the suppression of direct photon production induced by our anomalous quark-photon vertex in the QGP.

\section{Acknowledgments}

We are grateful to Aleksander Dorokhov, Sergo Gerasimov and Michael Ivanov for the useful discussions.
 This work was partially supported by the National Natural
 Science Foundation of China (Grant No. 11575254 and 11175215), and by the Chinese Academy of Sciences visiting
professorship for senior international scientists (Grant No.
2013T2J0011) and President's international fellowship initiative (Grant No. 2017VMA0045).


\end{document}